\documentclass[aps,pra,superscriptaddress,twocolumn]{revtex4-1}

\usepackage{bm}
\usepackage{graphicx}
\usepackage{longtable}
\usepackage{amssymb,amsmath,amsfonts,amsthm}

\newcommand{\ket}[1]{|{#1}\rangle}
\newcommand{\bra}[1]{\langle{#1}|}
\newcommand{\exval}[1]{\langle{#1}\rangle}
\newcommand{\bexval}[1]{\big\langle{#1}\big\rangle}

\newcommand{\fident}{\mathcal{F}_{{\rm ent}}}
\newcommand{\fidbell}{\mathcal{F}_{\rm ent}^{\rm Bell}}
\newcommand{\concbell}{\mathcal{C}_{\scriptscriptstyle{\rm CB}}^{\rm Bell}}

\newcommand{\rA}{{\rm{A}}}
\newcommand{\rB}{{\rm{B}}}
\newcommand{\rC}{{\rm{C}}}
\newcommand{\id}{\openone}

\DeclareMathOperator*{\Tr}{Tr}
\renewcommand\Re{\operatorname{Re}}
\renewcommand\Im{\operatorname{Im}}

\begin{document}
\title{Long quantum channels for high-quality entanglement transfer}

\author{L. Banchi}
\affiliation{Dipartimento di Fisica, Universit\`a di Firenze,
             Via G. Sansone 1, I-50019 Sesto Fiorentino (FI), Italy}
\affiliation{INFN Sezione di Firenze, via G.Sansone 1,
             I-50019 Sesto Fiorentino (FI), Italy}

\author{T.~J.~G.  Apollaro}
\affiliation{Dipartimento di Fisica, Universit\`a di Firenze,
             Via G. Sansone 1, I-50019 Sesto Fiorentino (FI), Italy}

\author{A. Cuccoli}
\affiliation{Dipartimento di Fisica, Universit\`a di Firenze,
             Via G. Sansone 1, I-50019 Sesto Fiorentino (FI), Italy}

\affiliation{INFN Sezione di Firenze, via G.Sansone 1,
             I-50019 Sesto Fiorentino (FI), Italy}

\author{R. Vaia}
\affiliation{Istituto dei Sistemi Complessi, Consiglio Nazionale delle Ricerche,
             via Madonna del Piano 10, I-50019 Sesto Fiorentino (FI), Italy}

\author{P. Verrucchi}
\affiliation{Istituto dei Sistemi Complessi, Consiglio Nazionale delle Ricerche,
             via Madonna del Piano 10, I-50019 Sesto Fiorentino (FI), Italy}
\affiliation{Dipartimento di Fisica, Universit\`a di Firenze,
             Via G. Sansone 1, I-50019 Sesto Fiorentino (FI), Italy}
\affiliation{INFN Sezione di Firenze, via G.Sansone 1,
             I-50019 Sesto Fiorentino (FI), Italy}
\date{\today}

\begin{abstract}
High-quality quantum-state and entanglement transfer can be achieved
in an unmodulated spin bus operating in the ballistic regime, which
occurs when the endpoint qubits A and B are coupled to the chain by
an exchange interaction $j_0$ comparable with the intrachain
exchange. Indeed, the transition amplitude characterizing the
transfer quality exhibits a maximum for a finite optimal value
$j_0^{\rm{opt}}(N)$, where $N$ is the channel length. We show that
$j_0^{\rm{opt}}(N)$ scales as $N^{-1/6}$ for large $N$ and that it
ensures a high-quality entanglement transfer even in the limit of
arbitrarily long channels, almost independently of the channel
initialization. For instance, the average quantum-state transmission
fidelity exceeds 90\,\% for any chain length. We emphasize that,
taking the reverse point of view, should $j_0$ be experimentally
constrained, high-quality transfer can still be obtained by adjusting
the channel length to its optimal value.
\end{abstract}

\maketitle

\section{Introduction}

Quantum state transfer between distant qubits (say, A and B) is a
fundamental tool for processing quantum information. The task
of covering relatively large distances between the elements of a
quantum computer, much larger than the qubit interaction range, can
be achieved by means of a suitable communication channel connecting
the qubits A and B.

Spin chains are among the most studied channel
prototypes~\cite{Bose2007}. In particular, the $XY$ $S\,{=}\,1/2$
model has been widely employed as a tool for testing and analyzing
quantum communication protocols. For such a communication channel to
be experimentally feasible one could be lead to consider systems with
uniform
intrachannel interactions~\cite{Bose2003,WojcikLKGGB2005,%
ApollaroP2006,CamposVenutiGIZ2007,BACVV2010,%
RamanathanCVC2011,ZwickO2011} and whose operation does not require
peculiar initialization procedures, although other proposals have
been recently put
forward~\cite{ChristandlDDEKL2005,KarbachS2005,DiFrancoPK2008,ZwickASO2011}.
On the other hand, the quantum-transfer capabilities of homogeneous
spin channels have not yet been fully explored under many respects.
For instance, it is a common belief that the longer the chain the
worse is the transmission
fidelity~\cite{Bose2007,Bose2003,HuL2009,FeldmanKZ2010} as an effect
of dispersion, and chains up to only a few tenths of spins are often
considered.

The present work is devoted to the study of the ballistic regime,
where the transmission can be
depicted~\cite{OsborneL2004,BACVV2010,Yadsan-ApplebyO2011} in terms
of a traveling wavepacket carrying the information about the state of
the endpoint qubit A, eventually yielding the state reconstruction at
the opposite endpoint qubit B thanks to the overall system's mirror
symmetry~\cite{ChristandlDDEKL2005,KarbachS2005,YungB2005,DiFrancoPK2008}.
The ballistic regime differs from that arising in the limit of very
weak endpoint
couplings~\cite{WojcikLKGGB2005,CamposVenutiGIZ2007,GualdiKMT2008,YaoJGGZDL2011}
where (almost) perfect state transfer occurs at very long times as a
result of a Rabi-like population transfer involving only the
lowest-energy single-particle modes. Understanding the basic
mechanism of ballistic transfer, where the number of involved
single-particle modes will be shown to be of the order of $N^{2/3}$,
allows us to devise an optimal value of the endpoint interactions for
any $N$, and vice versa. Remarkably, the corresponding transmission
quality, as witnessed by the state- and the entanglement fidelity,
does not decrease to zero when the channel becomes very long, but
remains surprisingly high.

\begin{figure}[t]
\includegraphics[width=80mm,angle=0]{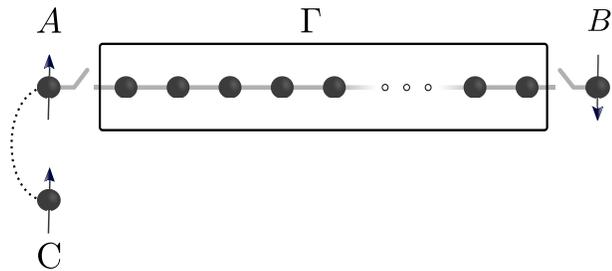}
\caption{The endpoints of a quantum channel $\Gamma$ are coupled to
the qubits A and B, via a tunable interaction $j_0$; A can be
entangled with an external qubit C. }
\label{f.qwire}
\end{figure}

We consider the setup illustrated in Fig.~\ref{f.qwire}: the channel
connecting the qubits A and B is a one-dimensional array of $N$
localized $S=1/2$ spins with exchange
interactions of $XX$ Heisenberg type and a possible external magnetic
field applied along the $z$ direction. This gives the total
Hamiltonian the following structure
\begin{eqnarray}
 \mathcal{H}&=& -\sum_{i=1}^{N-1}\big(S_i^x S_{i+1}^x+S_i^y S_{i+1}^y\big)
  - h\sum_{i=1}^{N} S_i^z
\label{e.Htot}
\\ \notag
 &&
 {-}j_0\!\!\sum_{i=0,N}\big(S_i^x S_{i+1}^x{+}S_i^y S_{i+1}^y\big)
 {-}h_0\big(S_0^z{+}S_{N+1}^z\big)~,
\end{eqnarray}
where the qubits A and B sit at the endpoint sites $0$ and $N{+}1$ of
a one-dimensional discrete lattice on whose sites $1,2,...,N$ the
spin chain is set. The exchange interaction (chosen as energy unit) and the
magnetic field $h$ are homogeneous along the chain, and an overall
mirror symmetry is assumed, implying the endpoint coupling $j_0$ and
field $h_0$ to be the same for both ends. The $N$ spins
constituting the $XX$ channel are collectively indicated by $\Gamma$.
We will focus our attention on how the state of the qubit B evolves
under the influence of the chain $\Gamma$, and depending on the
initial state of the qubit A; the latter is possibly entangled with
an ancillary qubit C. The results of the analysis are used for
gathering insights on the quantum-information transmission through
the chain, so as to characterize the dynamical evolution of the
overall system and to maximize the quality of the quantum-state
transfer.

Even though the overall scheme could also be used for realizing tasks
other than quantum information transfer, via the dynamical
correlations that the chain induces between A and B~\cite{BBVB2010},
our approach is specifically tailored for studying transfer processes
along the chain: the qubit B or the qubit-pair BC are considered as
target system, depending on whether the quantum-state of the
qubit A or that of the qubit-pair AC are to be transferred,
respectively.

In Section~\ref{s.dynqual} the formalism used to study the dynamical
evolution of the composite system is introduced and we derive the
corresponding time-dependent expressions for the quantities used for
estimating the quality of the quantum-state and entanglement transfer
processes. In Section~\ref{s.xxmodel} the proposed formalism is
applied to the $XX$ model described by Eq.~\eqref{e.Htot}. In
Section~\ref{s.optimal} we put forward an analytical framework in
order to improve the understanding of the conditions inducing an
optimal ballistic dynamics. The resulting high-quality transfer
processes along the spin chain are analyzed in
Section~\ref{s.exploit}. Conclusions are drawn in
Section~\ref{s.conclusions}, where comments about possible
implementation of the procedure are also put forward. Relevant
details of calculations are reported in the Appendices.

\section{Dynamics and transfer quality}
\label{s.dynqual}

\subsection{Dynamics}
\label{ss.dynamics}

The density matrix of a qubit B can be written as
$\rho=\sum_{\mu}b_{\mu}\,\zeta_{\mu}$ where $\mu=1,...,4$, $b_\mu$
are complex numbers, and $\{\zeta_\mu\}$ is an orthonormal basis in
the Hilbert space of $2\,{\times}\,2$ matrices endowed with the
Hilbert-Schmidt trace-product,
$\Tr[\zeta_\mu^\dagger\zeta_\nu]\,{=}\,\delta_{\mu\nu}$. In some
cases it is useful to choose Hermitian $\zeta_\mu$, for instance the
identity and the three Pauli matrices. However, a more direct
connection with the computational basis
$\{\ket{0}\equiv\ket{\uparrow},~\ket{1}\equiv\ket{\downarrow}\}$ for
the pure states of B is established by choosing
$\zeta_\mu=\ket{i}\bra{j}$ and setting $\mu=1\,{+}2i\,{+}\,j$, with
$i,\,j\,{=}\,0,\,1$. In what follows we will use this description,
and hence use indices $\mu,\,\nu,\,\lambda$ running from 1 to 4,
understanding summation over any repeated index, e.g.,
$\rho=b_{\mu}\,\zeta_{\mu}$. Hermiticity of $\rho$ implies $b_1$ and
$b_4$ to be real and $b_2\,{=}\,b_3^*$, while $\Tr\rho\,{=}\,1$ means
$b_1\,{+}\,b_4=1$, so that only three real parameters are
independent. The positivity of $\rho$ requires that
$|b_2|^2\,{\le}\,b_1b_4$.

The dynamics of B is described by its time-dependent density operator
$\rho(t)$; this can always be expressed as a linear function of a
suitable input density matrix $\rho^{\rm{in}}$, which may be the
initial one of qubit B itself or that of any other qubit playing a
role in its evolution:
\begin{equation}
 \rho(t)={\cal{E}}_t~\rho^{\rm{in}}~,
\label{e.rho(t)}
\end{equation}
where the quantum operation ${\cal{E}}_t$ is a trace-preserving,
completely positive, convex-linear map from {\em input} density
operators $\rho^{\rm{in}}$ to density operators of the qubit at time
$t$ (see, e.g., Ref.~\onlinecite{NielsenC2000}). The effect of the
linear map can be represented in terms of a $4\,{\times}\,4$
time-dependent matrix
\begin{equation}
 T_{\mu\nu}(t)= \Tr \big[\,\zeta_\mu^\dagger~{\cal{E}}_t~\zeta_\nu\big]~.
\label{e.T_munu}
\end{equation}
In fact, by writing $\rho(t)\,{=}\,b_\mu(t)~\zeta_\mu$ and
$\rho^{\rm{in}}=a_\mu\zeta_\mu$, one gets
\begin{equation}
 b_\mu(t) = T_{\mu\nu}(t)~ a_\nu~.
\label{e.rho(t)withT}
\end{equation}
If B is part of a larger system S ruled by a total Hamiltonian
${\cal{H}}$ and prepared in the initial state $\rho^{\rm{tot}}$, it
is also
\begin{equation}
 b_\mu(t)=\bexval{\zeta_\mu^\dagger(t)}~,
\label{e.rho(t)withAve}
\end{equation}
where $\exval{\,\cdot\,}\equiv\Tr[\,\cdot~\rho^{\rm{tot}}]$ and
$\zeta_\mu(t)=e^{\imath{\cal{H}}t}\,\id_{\rm{S}\smallsetminus{\rm{B}}}
\,{\otimes}\,\zeta_\mu\,e^{-\imath{\cal{H}}t}$, $\zeta_\mu$ acting on the
Hilbert space of the qubit B. The matrix elements $T_{\mu\nu}(t)$
explicitly follow from Eq.~\eqref{e.rho(t)withT}
and~\eqref{e.rho(t)withAve}.

Here we will essentially focus on the time evolution of the qubit B
and take the input density operator in Eq.~\eqref{e.rho(t)} to
represent the initial state of the qubit A, as this is the most
natural setup for studying quantum state and entanglement
transmission processes from A to B. Referring to the specific setup
described in Fig.~\ref{f.qwire} we prepare the overall system
A$\cup\Gamma\cup$B in the initial state
$\rho^{\rm{tot}}=\rho^\rA\,{\otimes}\,\rho^{\Gamma}\,{\otimes}\,\rho^\rB$
and let it evolve into
$e^{-\imath{\cal{H}}t}\rho^{\rm{tot}}e^{\imath{\cal{H}}t}$, where
${\cal{H}}$ is the total Hamiltonian~\eqref{e.Htot}. Notice that
$\rho^{\rm{tot}}$ has a fully separable structure: this is not a
necessary condition for the explicit determination of the dynamical
matrix (a structure such as $\rho^\rA\otimes\rho^{\Gamma B}$ would
also work), but it is quite meaningful in our scheme, as the qubit B
is assumed not to interact with the chain during the initialization
process, which makes quite artificial its being entangled with the
chain before the dynamics starts.

The density matrix of B evolves according to $\rho^\rB(t)=\Tr_{{\rm
A}\cup\Gamma}
\big[ e^{-\imath{\cal{H}}t}\rho^{\rm{tot}}e^{\imath{\cal{H}}t}\big]$, from which
\begin{eqnarray}
 T_{\mu\nu}(t)&=& {\Tr}_\rB\Big[\zeta_\mu^\dagger\
  {\Tr}_{\rA\cup\Gamma}\big[ e^{-\imath{\cal{H}}t}
  \zeta_\nu\,{\otimes}\,\rho^\Gamma
   \,{\otimes}\,\rho^\rB \,e^{\imath{\cal{H}}t}\big]\Big]
\notag\\
  &=& \bexval{\id^\rA\,{\otimes}\,\id^\Gamma
   \,{\otimes}\,\zeta_\mu^\dagger(t)}_\nu
\label{e.Texpl}
\end{eqnarray}
where
$\exval{~\cdot~}_\nu\equiv
\Tr[\,\cdot~\zeta_\nu\,{\otimes}\,\rho^\Gamma\,{\otimes}\,\rho^\rB]$.
Notice that the initial state of the qubit A does not enter the
expression of $T_{\mu\nu}(t)$, setting this procedure in the
framework of general tomographic approaches.

Let us now consider the time evolution of the quantum state
describing the qubit-pair $\rC{\cup}\rB$ when the total system is
initially prepared in the state
$\rho^{\rm{tot}}=\rho^{\rC\rA}\,{\otimes}\,\rho^\Gamma\,{\otimes}\,\rho^\rB$.
By a trivial generalization of the description introduced at the
beginning of this section, we write
$\rho^{\rC\rA}=g^{\rC\rA}_{\mu\nu}\,\zeta_\mu\,{\otimes}\,\zeta_\nu$,
and find
\begin{equation}
 \rho^{\rC\rB}(t)= \Big[\,g^{\rC\rA}_{\mu\lambda}\,T_{\nu\lambda}(t)\Big]
  \zeta_\mu\,{\otimes}\,\zeta_\nu
 \equiv g^{\rC\rB}_{\mu\nu}(t)~\zeta_\mu\,{\otimes}\,\zeta_\nu~.
\label{e.rhoCBt}
\end{equation}
This equation shows that $T_{\mu\nu}(t)$ is the only ingredient
needed not only for deriving the dynamics of the qubit B, but also of
the qubit pair $\rC{\cup}\rB$, providing the pair $\rC{\cup}\rA$ was
prepared separately from the rest of the system, and C is
non-interacting~\cite{DiFrancoPPK07}. In particular, if C and A are initially prepared in
one of the Bell states, say $(\ket{00}+\ket{11})/\sqrt{2}$, it is
$g^{\rC\rA}_{\mu\nu}=\frac{1}{2}\delta_{\mu\nu}$ (similar conditions
hold for the other Bell states), i.e.,
\begin{equation}
 \rho_{\rm{Bell}}^{\rC\rA}=\frac12~\zeta_\mu\otimes\zeta_\mu~,
\label{e.rhoCAbell}
\end{equation}
and hence
\begin{equation}
 \rho_{\rm Bell}^{\rC\rB}(t)
  ={\textstyle\frac12}\,T_{\nu\mu}(t)\,\zeta_\mu\,{\otimes}\,\zeta_\nu~.
\label{e.rhoCBbell}
\end{equation}

The trace-preserving property of the dynamical map,
$\Tr\big[{\cal{E}}_t\,\zeta_\mu\big]\,{=}\,\Tr\,\zeta_\mu$, together
with the specific choice of the basis $\{\zeta_\mu\}$, entails
$T_{12}\,{+}\,T_{42}\,{=}\,T_{13}\,{+}\,T_{43}\,{=}\,0$ and
$T_{11}\,{+}\,T_{41}\,{=}\,T_{14}\,{+}\,T_{44}\,{=}\,1$; the
Hermiticity of $\rho^\rB$ further implies that
$T_{\mu{1}}a_\mu$ and $T_{\mu{4}}a_\mu$ are real and
$T_{\mu{2}}a_\mu\,{=}\,T_{\mu{3}}^*a^*_\mu$.
Moreover, from Eq.~\eqref{e.Texpl} it clearly follows that symmetry
properties of the total Hamiltonian and of the initial state
$\rho^{\Gamma}\,{\otimes}\,\rho^\rB$ may result in constraints for
the matrix elements $T_{\mu\nu}$. In particular, when both are
symmetric under rotations around the $z$-axis, it is
\begin{equation}
 T_{\mu{2}}=T_{2\mu}=T_{22}~\delta_{\mu{2}}
~,~~~
 T_{\mu{3}}=T_{3\mu}=T_{33}~\delta_{\mu{3}}~,
\label{e.XXZcondition}
\end{equation}
and only three matrix elements, say $T_{11}$, $T_{22}$, and $T_{44}$,
need to be determined.

\subsection{Quality of transfer processes}
\label{ss.quality}

In order to study the quality of the transfer processes mediated by
the spin chain, we specifically consider the entanglement transfer
from C${\cup}$A to C${\cup}$B, when the former spin pair is initially
prepared in a pure state. In this case, the entanglement fidelity, which
is a proper measure of the quality of entanglement transmission, reads
$\fident(t)\equiv\Tr\big[{\rho^{\rC\rA}}^\dagger\rho^{\rC\rB}(t)\big]$,
which is, via Eq.~\eqref{e.rhoCBt},
\begin{equation}
 \fident(t)=T_{\mu\nu}(t)\,
 ~\left({g^{\rC\rA}_{\lambda\mu}}\right)^*\,g^{\rC\rA}_{\lambda\nu}~.
\label{e.entfid}
\end{equation}
The entanglement fidelity measures how close the state of C${\cup}$B
at time $t$ is to the initial state of C${\cup}$A. In particular, if
C and A are initially prepared in a Bell state,
Eq.~\eqref{e.rhoCAbell}, it is
\begin{equation}
 \fidbell(t)= {\textstyle\frac14}\, T_{\mu\mu}(t)~,
\label{e.fidbell}
\end{equation}
while the entanglement of the pair C${\cup}$B is measured by
\begin{equation}
 \concbell(t)=
 {\textstyle\frac12}\,
 \mathcal{C}\big[T_{\nu\mu}(t)\,\zeta_\mu\,{\otimes}\,\zeta_\nu\big]~,
\label{e.concbell}
\end{equation}
where $\mathcal{C}[\rho]$ is the concurrence~\cite{Wootters1998}
between two qubits in the state $\rho$. Note that
Eqs.~\eqref{e.rhoCBbell}, \eqref{e.fidbell}, and~\eqref{e.concbell}
are independent of which Bell state is chosen, since different Bell
states are connected to $\left(\ket{00}+\ket{11}\right)/\sqrt{2}$ by
unitary operations on C, which is isolated.

Another relevant tool for evaluating the quality of the transfer
processes is the fidelity of transmission from A to B, which reads
$\mathcal{F}_{_{\rA\rB}}(t)\equiv\Tr\big[{\rho^\rA}^\dagger\rho^\rB(t)\big]$
provided that ${\rho^\rA}$ is a pure state, i.e., by
Eq.~\eqref{e.rho(t)withT},
\begin{equation}\label{e.fid}
 \mathcal{F}_{_{\rA\rB}}(t)
  = b^{}_\mu(t)\,a^*_\mu =T_{\mu\nu}(t)~a^*_\mu a^{}_\nu~.
\end{equation}
If A is initially prepared in the state
$\ket{\psi_{\theta\varphi}}=
\cos\frac\theta{2}\ket{0}+e^{\imath\varphi}\sin\frac\theta{2}\ket{1}$,
meaning $a_1=\cos^2\frac\theta{2}$, $a_4=\sin^2\frac\theta{2}$, and
$a_2=a^*_3=e^{\imath\varphi}\sin\frac\theta{2}\cos\frac\theta{2}$,
Eq.~\eqref{e.fid} can be averaged over all possible initial pure
states by integrating over the Bloch sphere, resulting in
\begin{equation}\label{e.avefid1}
 \overline{\mathcal{F}}_{_{\rA\rB}}(t)
 ={\textstyle\frac13}+{\textstyle\frac16}\,T_{\mu\mu}(t)\big]~,
\end{equation}
which can be compared with Eq.~\eqref{e.fidbell} to obtain the
relation~\cite{Nielsen2002,HorodeckiHH1999}
\begin{equation}\label{e.avefid2}
 \overline{\mathcal{F}}_{_{\rA\rB}}(t)
  ={\textstyle\frac13}+{\textstyle\frac23}\,\fidbell(t) ~.
\end{equation}
It is worth noticing that a high average fidelity could still allow
for states that are poorly (or even not at all) transferred, while
the ultimate goal is the transmission of \emph{any} state.

When the setup is such that Eqs.~\eqref{e.XXZcondition} hold, from
Eqs.~\eqref{e.fidbell} and~\eqref{e.concbell} it follows that
\begin{eqnarray}
 \fidbell(t)&=&{\textstyle\frac14}\big(T_{11}+T_{44}+2\Re T_{22}\big) ~,
\\
 \concbell(t)&=&\max\big\{0,\,|T_{22}|-\sqrt{T_{14}T_{41}}\big\}~.
\label{e.concbellXXZ}
\end{eqnarray}
Moreover, if A is prepared in the pure state
$\ket{\psi_{\theta\varphi}}$, Eq.~\eqref{e.fid} explicitly reads
\begin{eqnarray}
 && \mathcal{F}_{_{\rA\rB}}(t)=
 T_{11}\cos^4{\textstyle\frac\theta2}+T_{44}\sin^4{\textstyle\frac\theta2}
\notag\\
 &&~~~~
 +\big[2(1+\Re T_{22})-(T_{11}+T_{44})\big]
 \sin^2{\textstyle\frac\theta2}\cos^2{\textstyle\frac\theta2}~,~~
\label{e.fidXX2}
\end{eqnarray}
which shows that $f_0\,{\equiv}\,T_{11}$ and $f_1\,{\equiv}\,T_{44}$
are the transmission fidelities of the states
$\ket{0}\,{=}\,\ket{\psi_{0\varphi}}$ and
$\ket{1}\,{=}\,\ket{\psi_{\pi\varphi}}$, respectively, while
$f\,{=}\,\frac12(1{+}\Re T_{22})$ represents the transmission
fidelity of the states
$\big(\ket{0}+e^{\imath\varphi}\ket{1}\big)/\sqrt{2}
\,{=}\,\ket{\psi_{\frac\pi2\varphi}}$.
The fidelity~\eqref{e.fidXX2} only depends on
$\theta\,{\in}\,[0,\pi]$ and its extrema can be easily determined. To
this purpose rewrite Eq.~\eqref{e.fidXX2} as
\begin{equation}
 \mathcal{F}_{_{\rA\rB}}(t) =
   f + {\textstyle\frac12}(f_0{-}f_1)\cos\theta
     + {\textstyle\frac12}(f_0{+}f_1\,{-}\,2f)\cos^2\theta ~;~
\label{e.fidXX3}
\end{equation}
it follows that $\mathcal{F}_{_{\rA\rB}}(t)$ takes the values $f_0$
and $f_1$ at the extrema of the range $\cos\theta\,{\in}\,[-1,1]$ and
can have a minimum
\begin{equation}
 f_{\rm{m}} = f-\frac18\,\frac{(f_0\,{-}\,f_1)^2}{f_0{+}f_1{-}2f}
\end{equation}
in between provided that
$c_{\rm{m}}\,{=}\,\frac{f_1{-}f_0}{2(f_1{+}f_0{-}2f)}\in(-1,1)$.
Therefore the bounds of $\mathcal{F}_{_{\rA\rB}}(t)$ are:
\begin{eqnarray}
 \mathcal{F}^{\rm{min}}_{_{\rA\rB}}(t)&=&\left\{
 \begin{array}{ll}
 \min\{f_1,f_0\}~~~~~& {\rm{if}}~~|c_{\rm{m}}|\,{\ge}\,1  \\
 f_{\rm{m}}          & {\rm{if}}~~|c_{\rm{m}}|\,{<}\,1 ~~,
 \end{array}\right.
\label{e.fidminxx}
\\
 \mathcal{F}^{\rm{max}}_{_{\rA\rB}}(t)&=&\max\{f_1,f_0\}~.
\end{eqnarray}
One can see that if $f_1\,{=}\,f_0$ the best transmitted states are
$\ket{0}$ and $\ket{1}$ which, on the other hand, are the best and
the worst transmitted ones (or viceversa) if
$|c_{\rm{m}}|\,{\ge}\,1$. This peculiar role of the computational
states is obviously a consequence of the assumed symmetry.

\section{The $XX$ model}
\label{s.xxmodel}

\subsection{Dynamical evolution}
\label{ss.XX1}

In this section we specifically consider the
Hamiltonian~\eqref{e.Htot}. The system A${\cup}\Gamma{\cup}$B is
prepared in the state
$\rho^{\rm{tot}}=\rho^\rA\,{\otimes}\,\rho^\Gamma\,{\otimes}\,\rho^\rB$
where $\rho_\Gamma$ is any state invariant under rotations around the
$z$-axis, and $\rho^\rB=b_1\zeta_1+b_4\zeta_4$: this choice fulfills
the requisite of $U(1)$ symmetry of
$\rho^{\Gamma}\,{\otimes}\,\rho^\rB$ leading to
Eq.~\eqref{e.XXZcondition}. Referring to the usual Jordan-Wigner
transformation, we cast Eq.~\eqref{e.Htot} in the fermionic quadratic
form
\begin{equation}
 \mathcal{H} = \sum_{i,j} c_i^{\dagger} \Omega_{ij} c^{}_j
             = \sum_n \omega_n \, c_n^\dagger c^{}_n~,
\label{e.XXdiag}
\end{equation}
where $\{c_i,c_i^{\dagger}\}$ are fermionic operators whose
nearest-neighbor interaction is described by a
$(N{+}2)\,{\times}\,(N{+}2)$ tridiagonal mirror-symmetric matrix
\begin{equation}
\Omega = -\frac12
\begin{bmatrix}
 2h_0 &  j_0 &      &      &      &     &      \\
 j_0  &  2h  & 1    &      &      &     &      \\
      &  1   & 2h   & ~1~  &      &     &      \\
      &      &\ddots&\ddots&\ddots&     &      \\
      &      &      &  ~1~ & 2h   & 1   &      \\
      &      &      &      &  1   & 2h  &  j_0 \\
      &      &      &      &      & j_0 & 2h_0 \\
\end{bmatrix}~;
\label{e.Omega}
\end{equation}
an orthogonal transformation
${\cal{O}}\,{=}\,\{{\cal{}O}_{ni}\}$ diagonalizes $\mathcal{H}$ (see
Appendix~\ref{a.tridiag}) in terms of Fermi operators
$c_n\,{=}\,\sum_i{\cal{}O}_{ni}\,c_i$ and $c^\dagger_n$ which
annihilate/create excitations of energy
$\omega_n$~\cite{LiebSM1961,CozziniGZ2007}. The trivial
time-evolution of the $c_n$'s entails a time-dependent transformation
\begin{equation}
 c_i(t) = \sum_{i=0}^{N+1} U_{ij}(t)~c_j ~,
\label{e.bogotime}
\end{equation}
where
\begin{equation}
 U_{ij}(t)=\sum_{n}{\cal{O}}_{ni} {\cal{O}}_{nj}~e^{-\imath \omega_nt}~.
\label{e.bho}
\end{equation}
Reminding that $\zeta_{1,4}=\frac12(\id\pm\sigma^z)$,
$\zeta_2=\zeta_3^\dagger=\sigma^+$, and that $\zeta_\mu$ in
Eq.~\eqref{e.Texpl} acts on the qubit B, and provided that
Eqs.~\eqref{e.XXZcondition} hold, we find
\begin{eqnarray}\label{k.matxxex}
 T_{11}(t)&=&|u(t)|^2 + v(t)\notag \\
 T_{44}(t)&=&1 - v(t)\notag \\
 T_{22}(t)&=&-p\,\exval{\sigma_{N{+}1}^z}\,u(t) e^{\imath\alpha(t)}
\end{eqnarray}
where
\begin{eqnarray}
 u(t)&=&|U_{N{+}1,0}(t)|~,
\label{e.u(t)}\\
 \alpha(t)&=&\arg[U_{N{+}1,0}(t)]
\label{e.alpha(t)}\\
 v(t)&=&|U_{N{+}1,N{+}1}(t)|^2\frac{\exval{\sigma_{N{+}1}^z}+1}{2}
 +C_{N{+}1}(t)~,~~
\label{e.v(t)}\\
 C_i(t)&=&\sum_{j,j'=1}^N U_{ij}^*(t)U^{}_{ij'}(t)
  \Tr \big[\rho^\Gamma c_j^\dagger c^{}_{j'}\big]~,
\label{e.cit}
\end{eqnarray}
with $p=\Tr[P\rho^\Gamma]$, and
$P=\exp\big(\imath\pi\sum_{i=1}^Nc_i^{\dagger}c_i\big)
\equiv\prod_{i=1}^N(-\sigma_i^z)$ is the chain parity operator,
which is a constant of motion. In particular, when $\rho^\Gamma$ is
the ground state of the chain,
$p\,{=}\,\textrm{sign}\big[\det\tilde{\Omega}\big]$~\cite{CozziniGZ2007},
where $\tilde{\Omega}$ is obtained from $\Omega$ by referring only to
the chain, i.e., deleting the first and last row and column. In our
case, $p\,{=}\,(-1)^{[(N\cos^{-1}\!h)/\pi]}$, where $[\,\cdot\,]$
denotes the integer part, meaning that $p$ equals $(-1)$ to the power
of the number of negative energy fermionic eigenmodes of $\Gamma$.
For $h\,{=}\,0$ it reduces to $p\,{=}\,(-1)^{[N/2]}$.

Once the above expressions are evaluated, the dynamics of B directly
follows from the initial state of A via Eq.~\eqref{e.rho(t)withT}.
Similarly, the dynamics of the qubit pair C${\cup}$B follows from the
initial state of the qubit pair C${\cup}$A, via Eq.~\eqref{e.rhoCBt}.
Moreover, the time evolution of the magnetization along the chain can
be straightforwardly obtained from the formalism described above:
\begin{equation}\label{e.szn}
 \bexval{\sigma^z_i(t)}=\big|U_{i0}(t)\big|^2
 \exval{\sigma_0^z}+\big|U_{i,N+1}(t)\big|^2
\exval{\sigma_{N+1}^z}+G_i(t)~,
\end{equation}
where $G_i(t) = 2 C_i(t)+ |U_{N{+}1,0}|^2+ |U_{N{+}1,N{+}1}|^2-1$.
The r.h.s. of Eq.~\eqref{e.szn} shows three distinct contributions
related with the dynamics of the excitations initially present in A,
B, and the chain $\Gamma$, respectively. For $h\,{=}\,0$ and $N$
even, $G_i(t)\,{=}\,0$, meaning that $\exval{\sigma^z_i(t)}$ are
solely determined by two traveling excitations, starting from the
edges of the chain. These excitations, because of the single-particle
nature of the Hamiltonian, do not scatter, which has great relevance
as far as the transport properties of the chain are concerned, as
discussed in Section~\ref{s.optimal}. On the other hand, when the
chain is initialized in the fully polarized state
$\bigotimes_{j=1}^N\ket{\downarrow}$ the contribution $C_i(t)$
vanishes and $G_i(t)$ has the only effect of redefining the range of
the magnetization during the dynamics.

\subsection{Fidelities and concurrence}
\label{ss.XX2}

We again consider the system ruled by the Hamiltonian~\eqref{e.Htot}
in the setup described above, so that Eqs.~\eqref{e.XXZcondition}
hold. Using Eq.~\eqref{k.matxxex} we find the entanglement fidelity
and the average transmission fidelity of pure states
\begin{eqnarray}
 \fidbell(t) &=& {\textstyle\frac14}
    +{\textstyle\frac14}u^2(t) - {\textstyle\frac12}
    p\cos\alpha\,\exval{\sigma_{N+1}^z}\, u(t),
\label{e.entfidbellxx}
\\
 \overline{\mathcal{F}}^{\,\rm pure}_{_{AB}}(t) &=&
 {\textstyle\frac12}+{\textstyle\frac16}u^2(t)
 -{\textstyle\frac13} p\cos\alpha\,\exval{\sigma_{N{+}1}^z}\, u(t),
\label{e.avefid3}
\end{eqnarray}
as well as  the concurrence,
\begin{equation}
 \concbell(t) = \max\big\{0,{\cal{C}}_0\big\}~,
\label{e.concxxh0}
\end{equation}
where
\begin{equation}
  {\cal{C}}_0 = \big|p\langle\sigma_{N+1}^z\rangle\big|\, u(t)
                -\sqrt{v(t)[1-u^2(t)-v(t)]}~.
\end{equation}
From the above formulas it appears that the choice of the initial
state $\rho^{\Gamma}\,{\otimes}\,\rho^\rB$ plays an important
role~\cite{BBBV2011}: in particular, in order to get the largest
concurrence it must be
\begin{equation}\label{e.condconc}
 p\langle\sigma_{N+1}^z\rangle=\pm 1~,
\end{equation}
meaning that $\rho^\Gamma$ is an eigenstate of $P$ and the qubit
B is initially in a polarized state,
$\rho^\rB\,{=}\,\zeta_1$ or $\rho^\rB\,{=}\,\zeta_4$; as
for the initial state of the channel, the choices range, for example,
from its ground state to a fully polarized state.
Such limitation in the choice of the initial state might be overcome
by applying a two-qubit encoding and decoding on states $\rho^\rA$ and
$\rho^\rB$, respectively~\cite{YaoJGGZDL2011, MarkiewiczW2009}.
Similarly, for the transmission fidelity the condition
\begin{equation}\label{e.condfid}
-p \langle\sigma_{N+1}^z\rangle\cos\alpha = 1
\end{equation}
must hold. Notice that the l.h.s. of Eq.~\eqref{e.condfid} follows
from the rotation around the $z$-axis undergone by the state during
the transmission and can be treated by choosing a proper magnetic
field~\cite{Bose2003} or the parity of $N$, as well as by applying a
counter-rotation on the qubit B~\cite{BBBV2011}.

The above analysis shows that, once condition~\eqref{e.condfid} is
fulfilled the quality of the state and entanglement transfer mainly
depends on $u(t)$ and increases with it; the residual dependence on
$v(t)$ suggests that the latter quantity need being minimized. In
Ref.~\cite{BBBV2011} it has been shown that $v(t)$ is exactly zero
provided that both $\Gamma$ and $\rB$ are initialized in the fully
polarized state. On the other hand, since $U_{ij}(t)$ entering
Eqs.~\eqref{e.u(t)} and~\eqref{e.v(t)} is a unitary matrix, if a
certain time $t^*$ exists such that $u(t^*)$ is very close to unity,
then $v(t^*)$ must be close to zero, no matter the initialization.
This situation is related with the optimal dynamics studied in
Ref.~\cite{BACVV2010}: such relation stands on the analytical ground
developed in the next Section.

\section{Optimal dynamics}
\label{s.optimal}

We now have the tools for determining the conditions for a dynamical
evolution that corresponds to the best quality of the transmission
processes. In Appendix~\ref{a.tridiag} the algebraic problem of
diagonalizing the $XX$ Hamiltonian~\eqref{e.XXdiag} in the case of
nonuniform mirror-symmetric endpoint interactions,
Eq.~\eqref{e.Omega} is analytically solved. The eigenvalues of
$\Omega$ can be written as
\begin{equation}
 \omega_k = -h-\cos{k} ~,
\label{e.omegak}
\end{equation}
in terms of the pseudo-wavevector $k$, which takes $N{+}2$ discrete
values $k_n$ in the interval $(0,\pi)$: from Eqs.~\eqref{e.phik}
and~\eqref{e.kn} it follows that these values obey
\begin{equation}
 k_n = \frac{\pi\,n+\varphi_{k_n}}{N{+}3}~,~~~~~ (n=1,\,...,\,N{+}2)~,
\label{e.x0.kn}
\end{equation}
with
\begin{equation}
 \varphi_k = 2k-2\,\cot^{-1}\!\Big(\frac{\cot{k}}{\Delta}\Big)
 ~~\in(-\pi,\pi) ~,
\label{e.x0.phik}
\end{equation}
\begin{equation}
 \Delta=\frac{j_0^2}{2-j_0^2}~,
\label{e.Delta}
\end{equation}
where we have set $h_0=h$. From the above equations it follows that
the $k$'s correspond to the equispaced values $\pi{n}/(N{+}3)$,
slightly shifted towards $\pi/2$ of a quantity which is smaller than
$\pi/(N{+}3)$, so that their order is preserved: therefore $k$ can be
used as an alternative index for $n$, understanding that it takes the
values $k_n$, as done in Eq.~\eqref{e.omegak}. According to the
conclusions of the previous section, we focus on the transition
amplitude, Eqs.~\eqref{e.bho} and~\eqref{e.u(t)}-\eqref{e.alpha(t)},
which explicitly reads:
\begin{equation}
U_{N+1,0}(t)\equiv u(t)e^{\imath\alpha(t)}=-\sum_n\rho(k_n)~
e^{\imath (\pi n-\omega_{k_n}t)}~,
\label{e.amplitude}
\end{equation}
where, after Eq.~\eqref{e.Parlett4}, it is
\begin{equation}
 \rho(k)= \frac1{N{+}3}~\frac{\Delta(1{+}\Delta)}{\Delta^2+\cot^2{k}}~,
\label{e.x0.Parlett}
\end{equation}
and mirror symmetry is exploited according to Eq.~\eqref{e.mirror}:
the transition amplitude above is a superposition of phase factors
with normalized weights, $\sum_n\rho(k_n)\,{=}\,1$, entailing
$|u(t)|\,{\leq}\,1$ with equality holding when all phases are equal.
The distribution $\rho(k)$ is peaked at $k\,{=}\,k_0\,{=}\,\pi/2$ and
its width is characterized by the parameter $\Delta$,
Eq.~\eqref{e.Delta}, so that the smaller $j_0$ the narrower
$\rho(k)$.

As $u(t)$ essentially measures the state-transfer quality,
the condition for maximizing it at some time $t^*$,
i.e. $u(t^*)\simeq{1}$, is that all phases
$\pi{n}\,{-}\,\omega_{k_n}t^*$ almost equal each other. Assume for a
moment that
the $k$'s be equispaced values, as in~\eqref{e.kx0y1}, and that the
dispersion relation be linear, $\omega_k=vk$: then
Eq.~\eqref{e.amplitude} would read
\begin{equation}
 u(t) = \left|\sum_n \rho(k_n)~e^{\imath\pi n(1-t/t^*)}\right|~,
\end{equation}
with $t^*{=}(N{+}3)/v$, so that
$u(t^*)\,{=}\,\sum_n\rho(k_n)\,{=}\,1$, i.e., all modes give a
coherent contribution and entail perfect transfer. On the other hand,
in our case $\omega_k$ is nonlinear in $k$, and the $k_n$ are not
equally spaced due to the phase shifts~\eqref{e.x0.phik} entering
Eq.~\eqref{e.x0.kn}, so generally the different modes undergo
dispersion and lose coherence.

\subsection{Transfer regimes}
The dependence of $\Delta$ upon $j_0$ reveals the possibility of
identifying different dynamical regimes, characterized by a
qualitatively different distribution $\rho(k)$,
Eq.~\eqref{e.x0.Parlett}, and hence, as for the transfer processes, a
different behavior of the transition amplitude $u(t)$. For extremely
small $j_0$ the distribution $\rho(k)$ can be so thin that (for even
$N$) only two opposite small eigenvalues come into play, say
differing by $\delta\omega$, and perfect transmission will be
attained at a large time $t=\pi(\delta\omega)^{-1}$ (for odd $N$
there is a third vanishing eigenvalue at $k\,{=}\,\pi/2$ and still
the two identical spacings $\delta\omega$ do matter). This is the
Rabi-like regime also mentioned in the Introduction.

A different regime is observed when $j_0$ is increased: a few
more eigenvalues come into play and it may occur, in a seemingly
random way, that their spacings be (almost) commensurate with each
other, i.e., they can be approximated as fractions with the same
denominator $K$, yielding phase coherence at
$t_{_K}\,{=}\,\pi{K}$.
By recording the maximum of $u(t)$ over a fixed large time interval $T$,
as $j_0$ is varied (see Ref.~\cite{WojcikLKGGB2005}), a rapid and
chaotic variation is observed. This regime is clearly useless for the
purpose of quantum communication.

\begin{figure}
\includegraphics[height=80mm,angle=90]{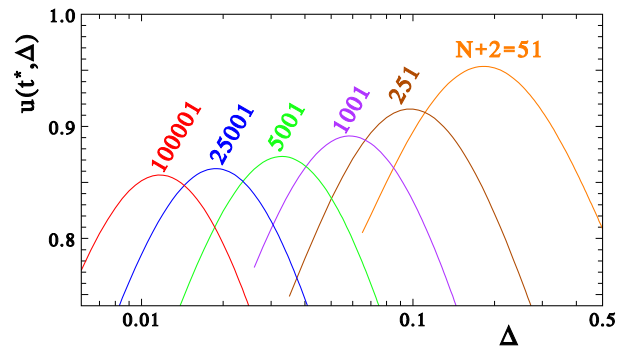}
\caption{(color online) Value of
$u(t^*,\Delta)$ as a function of $\Delta$, for different wire lengths
$N$. $t^*$ is obtained numerically by maximizing
Eq.~\eqref{e.amplitude} around $t\,{\simeq}\,N{+}3$.}
\label{f.utD}
\end{figure}

As $j_0$ further increases, the ballistic regime eventually manifests
itself: $\rho(k)$ involves so many modes that commensurability is
practically impossible, and a more regular behavior with short
transmission time $t^*\,{\sim}\,N$ sets in. The ballistic regime is
characterized by relatively large values of $u(t^*,\Delta)$ which is
the quantity plotted in Fig.~\ref{f.utD}, reporting numerical results
for increasing chain lengths. It appears that each curve shows a
maximum for a particular {\em optimal} value of
$\Delta\,{=}\,\Delta^{\rm{opt}}(N)$ or, equivalently, of
$j_0\,{=}\,j_0^{\rm{opt}}(N)$: such maxima are remarkably stable for
very high $N$ and yield very high transmission quality. In
Table~\ref{t.opt} we report some of the optimal values
$\Delta^{\rm{opt}}(N)$ and $j_0^{\rm{opt}}(N)$ for a wide interval of
chain lengths.
\begin{table*}
\begin{tabular}{|c|c|c|c|c|c|c|c|c|c|c|c|c|c|c|}
\hline
$N\,{+}\,2$ & 25&   51& 101&    251&    501&    1001&   2501&
   5001&   10001&  25001&  50001&  100001& 250001& 500001\\
\hline
\hline
$\Delta^{\rm opt}$ & 0.245&  0.182&  0.139&  0.098&  0.075&  0.058&
0.042&
  0.033&  0.026&  0.019&  0.015&  0.012&  0.009&  0.007\\
\hline
$j_0^{\rm opt}$ &
0.628&  0.556&  0.494&  0.422&  0.374&  0.332&  0.284&
  0.252&  0.224&  0.192&  0.171&  0.152&  0.130&  0.116\\
\hline
$u(t^*,\Delta^{\rm opt})$ & 0.972&  0.953&  0.936&  0.916& 0.902&
0.891&  0.880&
  0.873&  0.868&  0.862&  0.859&  0.857&  0.854&  0.853\\
\hline
\end{tabular}
\caption{Optimal values $\Delta^{\rm opt}$ and the corresponding
$j_0^{\rm opt}$ and $u(t^*,\Delta^{\rm opt})$ (see text for
details), for different $N$.}
\label{t.opt}
\end{table*}
This last `ballistic-transfer' regime is the one we are interested
in, since it has two strong advantages: first, the transmission time
$t^*\sim{N}$ is the shortest attainable, and second, the maximum
value $u(t^*,\Delta^{\rm opt})$ of $u(t^*,\Delta)$ is such that one
can achieve very good state transfer, e.g., the corresponding
transmission fidelity is far beyond the classical threshold, even for
very long chains.

The above analysis gives a physical interpretation of what is observed
in Fig.~3 of Ref.~\cite{WojcikLKGGB2005}, where the Rabi-like,
intermediate and ballistic regimes emerge.

\begin{figure}
\includegraphics[height=80mm,angle=90]{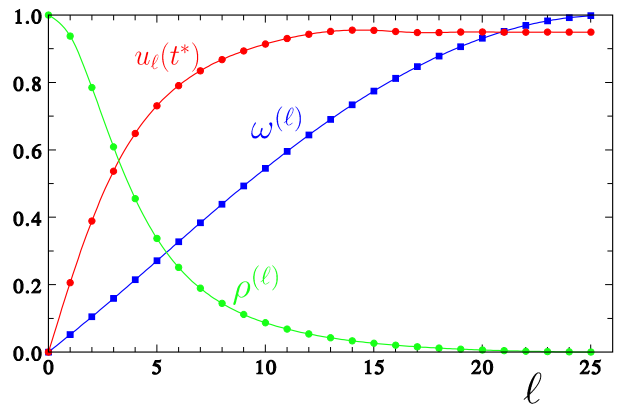}
\caption{(color online) Partial sum of the amplitude
$u_\ell(t^*)$ vs $\ell$ for $N\,{+}\,2\,{=}\,51$ and
$j_0\,{=}\,0.58$, together with the corresponding frequency and
density.}
\label{f.kcontrib}
\end{figure}

A qualitative picture of the ballistic regime can be obtained by
viewing the transition amplitude~\eqref{e.amplitude} as a wavepacket
with $N{+}2$ components. It can be evaluated by progressively adding
the contributions from symmetric eigenvalues, i.e., for odd $N$
summing between $(N{+}1)/2\mp{\ell}$, for
$\ell\,{=}\,0,\,1,\,...,\,(N{+}1)/2$. This yields the partial sum
$u_\ell(t^*)$ shown in Fig.~\ref{f.kcontrib}, together with the
corresponding frequency and density. One can see that the amplitude
increases only over the modes of the linear-frequency zone, i.e.
where frequencies are equally spaced, indicating that only those
wavepacket components whose frequency lies in such zone play a role
in the transmission process.

\subsection{Ballistic regime and optimal values}

\begin{figure}
\includegraphics[height=80mm,angle=90]{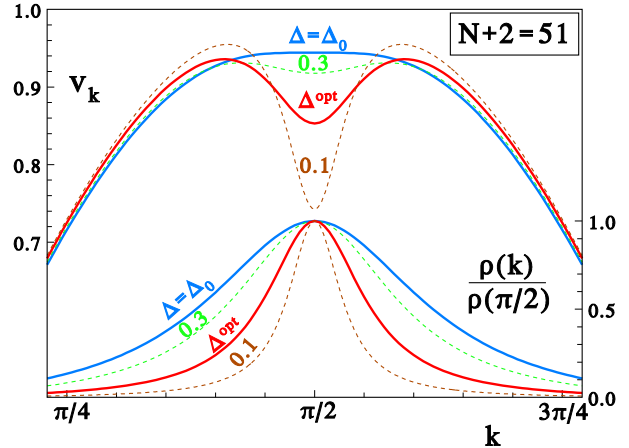}
\caption{(color online) `Group velocity'
$v_k\equiv[(N{+}3)/\pi]~\partial_n\omega_{k_n}$ and $\rho(k)$ vs $k$
for different values of $\Delta$. The thicker curves correspond to
$\Delta_0\,{=}\,0.3944$~\eqref{e.Delta0} that gives the flat behavior
at $k_0$, and to $\Delta^{\rm{opt}}\,{\simeq}\,0.1825$.}
\label{f.vg50}
\end{figure}

From the above reasoning, since the modes contributing to the
amplitude lie in a range of size $\Delta$ around $k_0$, in order to
get high-quality transfer processes it is necessary that the
corresponding frequencies be almost equally spaced, meaning that
$\omega_{k_n}$ is approximately linear in $n$. Actually, $\omega_k$
has an inflection point in $k_0$: its nonlinearity is of the third
order in $k\,{-}\,k_0$ and the modes close to $k_0$ satisfy the
required condition. However, from the phase-shifts~\eqref{e.x0.phik}
a further cubic term arises, which depends on $\Delta$. As $\Delta$
varies with $j_0$, the latter can be chosen so as to eliminate the
cubic terms, yielding a wide interval with almost constant frequency
spacing. The latter can be expressed just as the derivative of
$\omega_{k_n}$ with respect to $n$,
$\partial_n\omega_{k_n}=\sin{k}~\partial_nk$. The last term is
evaluated from Eqs.~\eqref{e.x0.kn} and~\eqref{e.x0.phik},
\begin{eqnarray}
 \partial_nk &=& \frac{\pi+\varphi_k'\partial_nk}{N{+}3}
  = \frac\pi{N{+}3-\varphi_k'}~,
\\
 \varphi_k'
 &=& -2\,\frac{1{-}\Delta}\Delta
  +\frac{2(1{-}\Delta^2)\cos^2k}{\Delta[\Delta^2+(1{-}\Delta^2)\cos^2k]} ~,
\label{e.x0.dphik}
\end{eqnarray}
so that
\begin{eqnarray}
 &&\partial_n\omega_{k_n} = \frac{\pi\,\sin{k}}{N{+}3-\varphi_k'}
\notag\\
 && ~~= \frac{\pi}{t^*}\,\Big[1+
   \Big(2\,\frac{1{-}\Delta^2}{t^*\Delta^3}{-}\frac12\Big)
   \cos^2\!{k}+O(\cos^4\!{k})\Big] \,,~~~~
\end{eqnarray}
where $t^*\,{=}\,N{+}3+2\,(1{-}\Delta)/\Delta$ is the arrival time.
It follows that one can minimize the nonlinearity of $\omega_{k_n}$
by setting the width to the value $\Delta_0$ satisfying
\begin{equation}
 \Delta_0 = \Big[\,\frac4{t^*}\,(1{-}\Delta_0^2)\Big]^{1/3}
 ~~\mathop{\longrightarrow}_{N{\gg}1}~~ 2^{2/3}N^{-1/3}~,
\label{e.Delta0}
\end{equation}
and $j_0\,{\simeq}\,2^{5/6}{N}^{-1/6}$ for large $N$. Therefore the
main mechanism that produces an optimal ballistic transmission is
that of varying the endpoint exchange parameter to the value $j_0$
that `linearizes' the dispersion relation. Actually, if the
corresponding $\Delta_0\,{=}\,\Delta(j_0)$ is such that $\rho(k)$
exceeds the region of linearity, further gain arises by lowering
$j_0$ so as to tighten the relevant modes towards $k_0$. However, at
the same time $\omega_{k_n}$ becomes less linear and the trade-off
between these two effects explains why a maximum is observed. This is
well apparent in Fig~\ref{f.vg50}, where for different values of
$\Delta$ the shapes of $\partial_n\omega_k$ can be compared with the
excitation density $\rho(k)$: for $\Delta\,{=}\,\Delta_0$ the density
still has important wings in the nonlinear zone, so the optimal value
$\Delta^{\rm{opt}}$ turns out to be smaller.

\begin{figure}\begin{center}
\includegraphics[width=.4\textwidth]{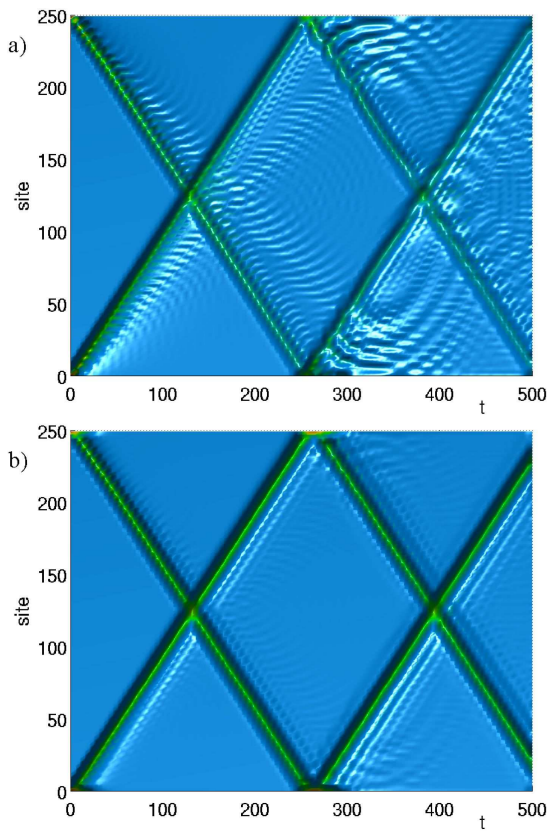}
\end{center}
\caption{(color online) Dynamics of the magnetization $\sigma_i^z(t)$
at time $t$ and site $i$ when a) $j_0\,{=}\,1$ and b)
$j_0\,{=}\,j_0^{\rm{opt}}$. The initial state of the whole system is
$\ket{\uparrow}\otimes\ket{\downarrow\downarrow\cdots\downarrow}\otimes
\ket{\uparrow}$ and the length of the chain is $N\,{+}\,2\,{=}\,250$.}
\label{f.sz}
\end{figure}

The dynamics in the ballistic regime is best illustrated by the time
evolution of the magnetization Eq.~\eqref{e.szn} along the chain,
plotted in Fig.~\ref{f.sz} when the initial state is
$\ket{\uparrow}\otimes
\ket{\downarrow\downarrow\cdots\downarrow}\otimes \ket{\uparrow}$.
The initial magnetizations at the endpoints generate two traveling
wavepackets: for non-optimal couplings ($j_0\,{=}\,1$, upper panel)
they change their shape and quickly straggle along the chain; for
optimal coupling ($j_0\,{=}\,j_0^{\rm opt}$, lower panel) they travel
with minimal dispersion. This confirms that the coherence is best
preserved when the optimal ballistic dynamics is induced: In the next
session we show that to such dynamics do in fact correspond high
values of the quality estimators for the state and entanglement
transfer.

\section{Information transmission exploiting optimal dynamics}
\label{s.exploit}
The requirement Eq.~\eqref{e.condfid}, means that the state is not
rotated by the dynamics when it arrives on site B, though during the
evolution it may undergo a rotation around the $z$ axis. In
Ref.~\cite{BBVB2010} it has been shown that
$\alpha\,{=}\,-\frac{\pi}2(N{+}1)$ at the transmission time $t^*$.
Therefore, also without applying a counter-rotation on qubit
B~\cite{BBBV2011}, condition~\eqref{e.condfid} can be fulfilled by
choosing $N\,{=}\,4M{\pm}1$ where the sign $\pm$ is given
by~\eqref{e.condconc} and thus depends on the initial state of the
chain. In the following we assume that conditions~\eqref{e.condconc}
and~\eqref{e.condfid} are always satisfied.

\begin{figure}
\includegraphics[height=80mm,angle=90]{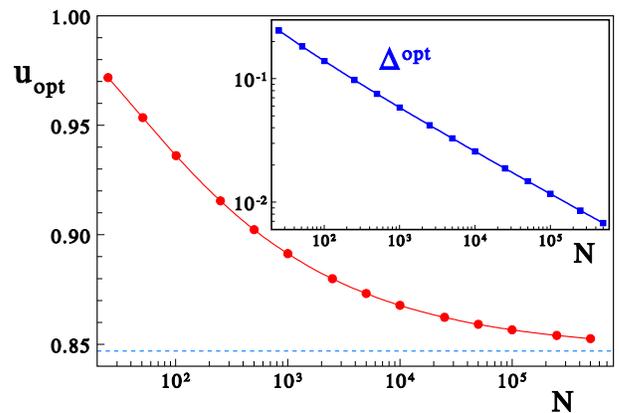}
\caption{(color online) Behavior of the maximum attainable amplitude
$u_{\rm{opt}}$ and (inset) of the corresponding optimal value of
$\Delta^{\rm{opt}}$ vs logarithm of the chain length $N$. The
horizontal dashed line is the infinite $N$ limit of $u_{\rm{opt}}$.}
\label{f.Umax}
\end{figure}

Let us consider for the moment that $\Gamma$ and B are initially in
the fully polarized state
$\ket{\downarrow\downarrow\cdots\downarrow}\otimes\ket{\downarrow}$.
In that case $v(t)\,{\equiv}\,0$ and the transmission
fidelities~\eqref{e.entfidbellxx} and~\eqref{e.avefid3}, as well as
the concurrence~\eqref{e.concxxh0}, only depend on, and monotonically
increase with, $u(t)$. The best attainable information transfer
quality corresponds therefore to the maximum amplitude
$u_{\rm{opt}}\,{\equiv}\,u(t^*,\Delta^{\rm{opt}})$. In
Fig.~\ref{f.Umax} and in Table~\ref{t.opt} we report these values
together with the corresponding optimal $\Delta^{\rm{opt}}$ as a
function of the chain length $N$ in a logarithmic scale; the inset
shows that $\Delta^{\rm{opt}}$ obeys the same power-law behavior
predicted in Eq.~\eqref{e.Delta0} for $\Delta_0$. Fig.~\ref{f.Umax}
also shows that for larger and larger $N$ the maximal amplitude
$u_{\rm{opt}}$ does not decrease towards zero, but it rather tends to
a constant value of about 0.85, which is surprisingly high, as, e.g.,
it corresponds to an average fidelity
$\overline{\mathcal{F}}_{_{\rA\rB}}(t^*)\,{\gtrsim}\,0.9$. This is
indeed true: We show in Appendix~\ref{a.largeN} that in the limit of
$N\,{\to}\,\infty$ the optimized amplitude tends to
$u_{\rm{opt}}\,{=}\,0.8469$. Basically, this tells us that it is
possible to transmit quantum states with very good quality also over
macroscopic distances. From Eq.~\eqref{e.scaling} we can derive the
asymptotic behavior of the optimal coupling
\begin{equation}
 j_0^{\rm{opt}} \simeq 1.030~N^{-1/6}~.
\end{equation}

\begin{figure}\begin{center}
\includegraphics[width=80mm]{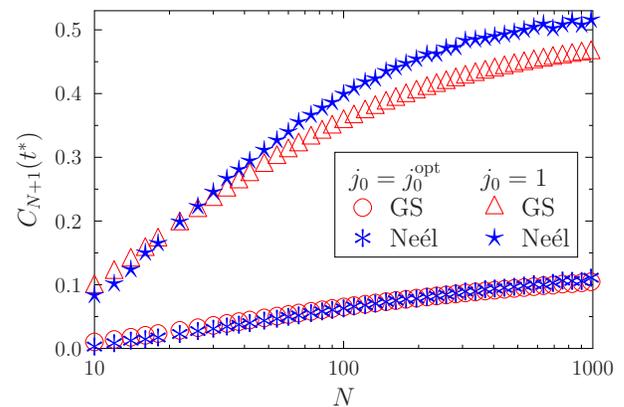}\end{center}
\caption{(color online) $C_{N{+}1}(t^*)$ for different initial states
of the chain (ground state, anti-ferromagnetic Ne\'el state, and
series of singlets~\cite{BBBV2011}) when $j_0\,{=}\,j_0^{\rm opt}$
and $j_0\,{=}\,1$. The results for a series of singlets are
numerically indistinguishable from those with the Ne\'el state. }
\label{f.ctstar}
\end{figure}

In the optimal ballistic case the channel initialization is not
crucial, as different initial states satisfying~\eqref{e.condconc}
give rise to almost the same dynamics as discussed at the end of
subsection III B. In fact, the term $C_{N{+}1}(t)$ entering
Eq.~\eqref{e.v(t)} essentially embodies the effect of channel
initialization and it is expected to be small at $t=t^*$. This is
apparent in Fig.~\ref{f.ctstar}, where for $j_0=j_0^{\rm opt}$,
$C_{N{+}1}(t^*)$ stays well below 0.1 for $N$ as long as 1000.
\begin{figure}\begin{center}
\includegraphics[angle=-90,width=80mm]{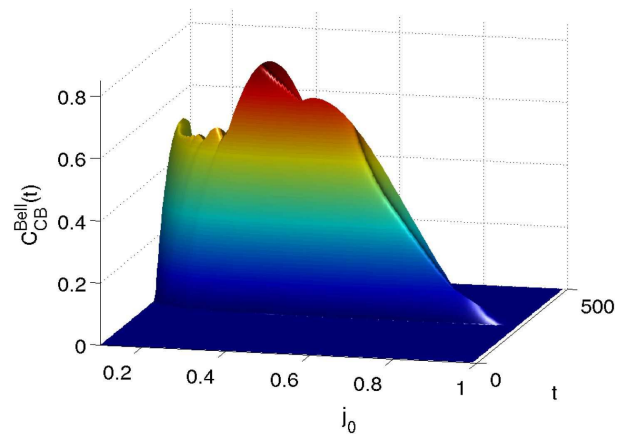}
\end{center}
\caption{(color online) Evolution of the concurrence $\mathcal{C}_{\rm
CB}^{\rm Bell}$ vs $j0$ and $t$. The length of the chain is
$N\,{+}\,2\,{=}\,250$.}
\label{f.cf}
\end{figure}

The transmitted entanglement, as measured by the
concurrence~\eqref{e.concxxh0}, is shown in Fig.~\ref{f.cf} as a
function of $j_0$ and $t$, with the channel initially prepared in its
ground state. As expected, the peak of the transmitted concurrence is
observed for $j_0=j_0^{\rm opt}$; away from $j_0^{\rm opt}$ the
quality of transmission falls down because $u(t^*)$ decreases and,
accordingly, $v(t^*)$ is allowed to increase. In fact, in the
non-optimal ballistic case the quality of entanglement transfer does
depend on the the initial state of the
channel~\cite{BayatB2010,BBBV2011}; for instance, when $j_0=1$ and
the chain is initially in its ground state, the contribution of the
overlap terms $\Tr[\rho^\Gamma{c^\dagger_j}c_{j'}]$ in
Eq.~\eqref{e.cit} is not quenched by the dynamical prefactors, and
higher values of $C_{N+1}(t^*)$ (see Fig.~\ref{f.ctstar}) inhibit the
transmission of entanglement even if $u(t^*)\neq 0$.
\begin{figure}\begin{center}
\includegraphics[width=.45\textwidth]{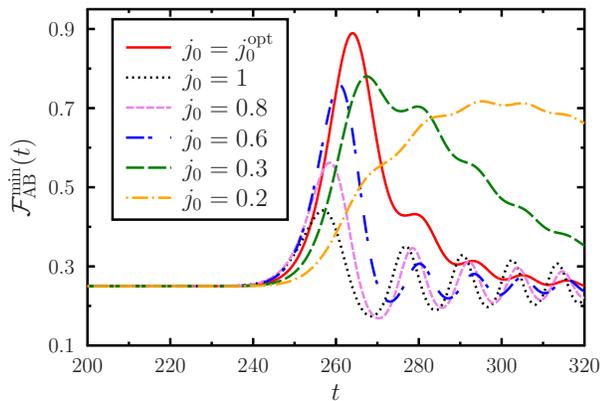}
\end{center}
\caption{(color online) Minimum fidelity vs time for different values
of $j_0$. The length of the chain is $N\,{+}\,2\,{=}\,251$ and
$j_0^{\rm{opt}}\,{=}\,0.422$.}
\label{f.minf}
\end{figure}

The effect of the optimization of $j_0$ is clearly evident in the
time behavior of the minimum fidelity, Eq.~\eqref{e.fidminxx},
reported in Fig.~\ref{f.minf}. The peak of
$\mathcal{F}^{\rm{min}}_{_{\rA\rB}}(t)$ for $j_0^{\rm{opt}}$ occurs
at the arrival time $N{+}3\,{+}\,s$ with a time delay $s$ that agrees
with the asymptotic value $s\,{\simeq}\,2.29\,N^{1/3}$ derived in
Appendix~\ref{a.largeN}. The `reading time', i.e., the time interval
during which the qubit B keeps being in the transferred quantum
state, is $t_{\rm{R}}\simeq\Delta^{-1}$, as the same figure also
shows; note that, in the optimal case, $t_{\rm{R}}$ increases with
$N$ according to the asymptotic behavior
$t_{\rm{R}}\,{\simeq}\,1.89\,N^{1/3}$.

\section{Conclusions}
\label{s.conclusions}

In this paper we have shown that high-quality quantum state and
entanglement transfer between two qubits A and B is obtained through
a uniform $XX$ channel of arbitrary length $N$ by a proper choice of
the interaction $j_0$ between the channel and the qubits. The value
of such interaction is found to control the transfer regime of the
channel, which varies, as $j_0$ increases, from the Rabi-like one,
characterized by very long transmission time, to an intermediate
regime, which turns useless for the purpose of quantum communication,
and finally becomes ballistic for $j_0$ of the order of the
intrachannel interaction.

In order to get coherent transfer in the ballistic regime, it is
desirable that the $k$-density of the traveling wavepacket generated
by Alice's initialized qubit A be narrow and concentrated in the
linear zone of the dispersion relation, i.e., with equispaced
frequencies. As the parameter $j_0$ controls both the width of the
$k$-density and the spacings of the frequencies entering the
dynamics, one can therefore improve the transmission quality up to a
best trade-off arising for an optimal value $j_{\rm{opt}}(N)$ which
for large $N$ behaves as $j_{\rm{opt}}(N)\,{\simeq}\,1.03\,N^{-1/6}$.
Remarkably, we have found that for such a choice the
quantum-state-transfer quality indicators are very high and, indeed,
have a lower bound for $N\,{\to}\,\infty$ that still allows to
efficiently perform quantum-information tasks: e.g., the average
fidelity of state transmission is larger than 90\,\%.

The ballistic regime ensures fast transmission on a time scale of the
order of $N$, at variance with the Rabi-like regime, and in the
optimal case the reading time increases as $N^{1/3}$. It is also to
be noted that, if experimental settings constrain to a given value
$j_0^{\rm{exp}}$, yet one can optimize the chain length in such a way
that $j_0^{\rm{exp}}=j_0^{\rm{opt}}(N)$. The only requirement on the
initial state of the receiving qubit B and the spin bus is to possess
$U(1)$ symmetry, a condition that can be fulfilled by several
configurations concerning the spin bus, ranging from the fully
polarized state to the highly-entangled ground state. If a large
magnetic field can be switched on during the initialization procedure
(in order to fully polarize the channel), and switched off as soon as
the transmission starts, then, from our analytical treatment it
emerges that temperature is not a major issue as far as the dynamical
evolution of the channel is concerned, though low temperatures are
obviously necessary to protect the qubits from phase and amplitude
damping due to the solid-state environment. To judge if the proposed
scheme identifies a reasonable experimental framework, let us
estimate the magnitude of the involved physical quantities. Consider
a solid-state implementation with lattice spacing of about {10~\AA}
and intrachain exchange $J\,{\simeq}\,10^2$~K. A quantum state will
then be transferred with fidelity 90\,\% along a channel of length
1~cm ($N\,{\simeq}\,10^7$) using
$j_0\,{\simeq}\,1.03\,N^{-1/6}\,J\,{\simeq}\,7.0$~K, with
transmission time
$t\,{=}\,N\,\hbar/(k_{\rm{B}}J)\,{\simeq}\,0.75\,\mu$s and reading
time
$t_{\rm{R}}\,{\simeq}\,1.9\,N^{1/3}\,\hbar/(k_{\rm{B}}J)\,{\simeq}\,0.03$\,ns.

\acknowledgements

We acknowledge the financial support of the Italian Ministry of
University in the framework of the 2008 PRIN program (contract N.
2008PARRTS 003). LB and PV gratefully thank Dr.~A.~Bayat and
Prof.~S.~Bose for useful discussions, and TJGA thanks ISC-CNR for the
kind hospitality.

\appendix

\section{Quasi-uniform tridiagonal matrices}
\label{a.tridiag}

The matrix $\Omega$, Eq.~\eqref{e.Omega}, can be written as
$\Omega=-h-\frac12{\cal{M}}$ where
\begin{equation}
{\cal M}(x,y) =
\begin{bmatrix}
 ~x~ &  y  &      &      &      &     &     \\
  y  & ~0~ &  1   &      &      &     &     \\
     &  1  & ~0~  & ~1~  &      &     &     \\
     &     &\ddots&\ddots&\ddots&     &     \\
     &     &      &  ~1~ & ~0~  &  1  &     \\
     &     &      &      &  1   & ~0~ &  y  \\
     &     &      &      &      &  y  & ~x~ \\
\end{bmatrix}
\label{e.MM}
\end{equation}
is a square tridiagonal matrix of dimension $M=N+2$,
and $x=2(h_0-h)$ and $y=j_0$.
This real symmetric matrix is diagonalized by an orthogonal matrix
$O(x,y)$,
\begin{equation}
 \sum_{i,j=1}^M {\cal O}_{ki} {\cal M}_{ij} {\cal O}_{k'\!j}
            = \lambda_k\,\delta_{kk'} ~,
\end{equation}
and it is known that
 ~(\emph{i})~ if $y\,{\ne}\,0$ the eigenvalues
are nondegenerate~\cite{Parlett1998},
 ~(\emph{ii})~ the eigenvectors corresponding to the eigenvalues
ordered in descending order are alternately symmetric and skew
symmetric~\cite{CantoniB1976}, i.e.,
\begin{equation}
 {\cal O}_{ki} = \pm~ {\cal O}_{k,M+1-i} ~.
\label{e.mirror}
\end{equation}

The eigenvalues are the roots of the associated characteristic
polynomial
\begin{equation}
 \chi_{_M}(\lambda;x,y) \equiv \det [\lambda-{\cal M}(x,y)] ~.
\label{e.chilambda}
\end{equation}

In the fully uniform case the characteristic polynomial is
$\eta_{_M}(\lambda)\equiv\chi_{_M}(\lambda;0,1)$ and one easily
obtains the recursion relation
\begin{equation}
 \eta_{_M}= \lambda\,\eta_{_{M-1}}-\,\eta_{_{M-2}}~,
 \label{e.etaeta}
\end{equation}
that can be solved in terms of Chebyshev polynomials of the second
kind,
\begin{equation}
  \eta_{_M} = \frac{\sin(M{+}1)k}{\sin{k}} ~,
\label{e.chebyshev}
\end{equation}
where
\begin{equation}
 \lambda \equiv 2\cos k ~,
\label{e.lambdak}
\end{equation}
so the eigenvalues of ${\cal{M}}(0,1)$ correspond to $M$ discrete
values of $k$,
\begin{equation}
 k = \frac{\pi\,n}{M{+}1}~~~~~~ [n=1, \dots, M]~;
\label{e.kx0y1}
\end{equation}
the corresponding eigenvectors are
\begin{equation}
 {\cal O}_{ki}(0,1) = {\textstyle\sqrt{\frac{2}{M{+}1}}~~\sin{ki}}~.
\label{e.Tx0y1}
\end{equation}

The general determinant~\eqref{e.chilambda} can be expressed in terms
of the $\eta_{_M}$'s by expanding it in the first and then in the
last column,
\begin{equation}
 \chi_{_M} = (\lambda^2{-}2x\lambda{+}x^2)\,\eta_{_{M-2}}
    {-}\,2y^2(\lambda{-}x)\,\eta_{_{M-3}} +y^4\,\eta_{_{M-4}}~,
\end{equation}
and using Eq.~\eqref{e.etaeta} one can eliminate the explicit
appearances of $\lambda$,
\begin{eqnarray}
 \chi_{_M} &=& \eta_{_M} -2x\,\eta_{_{M-1}}+x^2\,\eta_{_{M-2}}
\notag \\
 && +(1{-}y^2)\big[2\,\eta_{_{M-2}}{-}2x\,\eta_{_{M-3}}
 {+}(1{-}y^2)\,\eta_{_{M-4}}\big]~.
\label{e.chiM}
\end{eqnarray}
By rewriting Eq.~\eqref{e.chebyshev} as
~$\sin{k}~\eta_{_M}=\Im\big[e^{\imath(M{+}1)k}\big]$ and defining
\begin{equation}
 z^2 \equiv 1-y^2 ~,~~~~
 z_k^2 \equiv z^2\,e^{-2\imath k} ~,~~~~
 x_k \equiv x\,e^{-\imath k} ~,
\end{equation}
\begin{equation}
 u_k \equiv 1-x_k+z_k^2
     = e^{-\imath k}\big\{[(2{-}y^2)\cos{k}-x] +\imath\,y^2\sin{k}\big\}~,
\label{e.uk}
\end{equation}
Eq.~\eqref{e.chiM} takes the form
\begin{equation}
 \sin{k}~\chi_{_M}(k) =  \Im\big\{e^{\imath(M{+}1)k}~u_k^2 \big\}~.
\label{e.chik}
\end{equation}
The secular equation $\Im\big\{e^{\imath(M{+}1)k}~u_k^2 \big\}=0$
entails that when $k$ corresponds to an eigenvalue the quantity in
braces is real and equal to either $\pm|u_k|^2$; by Eq.~\eqref{e.uk}
it turns into $\sin\big[(M{+}1)k-\varphi_k\big]=0$, with the phase
shift
\begin{equation}
 \varphi_k =
 2k-2\tan^{-1}\!\frac{y^2\sin{k}}{(2{-}y^2)\cos{k}-x} ~,
\label{e.phik}
\end{equation}
so the $M$ eigenvalues correspond to
\begin{equation}
 k_n = \frac{\pi\,n+\varphi_{k_n}}{M{+}1}~,~~~~~ (n=1, \dots, M)~.
\label{e.kn}
\end{equation}

We are interested in the squared components of the first column of
the diagonalizing matrix ${\cal{O}}_{ki}$, which can be expressed
as~\cite{Parlett1998}
\begin{equation}
 {\cal O}_{k1}^2 = \frac{\xi_{_{M-1}}(\lambda_k)}
                 {\partial_\lambda\chi_{_M}(\lambda_k)}
          = -\frac{2\sin{k}~\xi_{_{M-1}}(k)}{\partial_k\chi_{_M}(k)}~,
\label{e.Parlett}
\end{equation}
where $k$ assumes the values~\eqref{e.kn} and
$\xi_{_{M-1}}(\lambda;x,y)$ is the characteristic polynomial
associated to the first minor matrix ${\cal{M}}_{11}$ that, expanded
in the last column and using Eq.~\eqref{e.etaeta}, reads
\begin{eqnarray}
 \xi_{_{M-1}} &\equiv& \det [\lambda-{\cal{M}}_{11}(x,y)]
\notag \\
 &=& (\lambda\,{-}\,x)\,\eta_{_{M-2}}-y^2\,\eta_{_{M-3}}
\notag \\
 &=& \eta_{_{M-1}}{-}\,x\,\eta_{_{M-2}}{+}\,(1{-}y^2)\,\eta_{_{M-3}} ~.
 \label{e.xieta}
\end{eqnarray}
Then the numerator of Eq.~\eqref{e.Parlett} is
\begin{equation}
 \sin{k}~\xi_{_{M-1}} = \Im\big\{e^{\imath Mk}u_k \big\}~,
\label{e.xiNm1}
\end{equation}
while from Eq.~\eqref{e.chik} one has
\begin{eqnarray}
 \sin{k}~\partial_k\chi_{_M}(k) &=&
      (M{+}1)\,\Re \big\{e^{\imath(M{+}1)k}~u_k^2 \big\}
\notag\\
    &&  +2\,\Im\big\{e^{\imath(M{+}1)k}~u_ku'_k\big\}~,
\end{eqnarray}
where the argument of $\Re$ is real indeed; retaining only the
dominant term for $M\,{\gg}\,1$ Eq.~\eqref{e.Parlett} becomes
\begin{equation}
 {\cal O}_{k1}^2 = \frac{2}{M{+}1} ~\frac{y^2\sin^2\!{k}}
 {[(2{-}y^2)\cos{k}-x]^2+y^4\sin^2\!{k}}~.
\label{e.Parlett4}
\end{equation}
For $x=0$ the above expression is in agreement with
Ref.~\cite{WojcikLKGGB2005}. In the most common case
$x\,{<}\,2{-}y^2$ the maximum $k_0$ of ${\cal{O}}_{k1}^2$ is located
at
\begin{equation}
 \cos{k_0} = \frac{x}{2{-}y^2}~,
\end{equation}
so that switching $x$ on the maximum shifts from $\pi/2$ to
\begin{equation}
 k_0 = \frac\pi{2}-\sin^{-1}\frac{x}{2{-}y^2}~;
\end{equation}
the `eigenvalue' corresponding to the maximum is
\begin{equation}
 \lambda_0 = 2\cos{k_0} = \frac{2x}{2{-}y^2}
\end{equation}
so that for $y\sim1$, the maximum shifts the `energy' linearly with
$x$. Expanding ${\cal{O}}_{k1}^2$ around the maximum, the leading
behavior is found to be a Lorentzian,
\begin{equation}
 {\cal O}_{k1}^2 \simeq \frac2{M{+}1}
 ~\frac{y^2}
 {y^4+[(2{-}y^2)^2-x^2](k-k_0)^2} ~,
\end{equation}
whose width (HWHM) is given by
\begin{equation}
 \Delta \simeq \frac{y^2}{\sqrt{(2{-}y^2)^2-x^2}} ~.
\end{equation}
When $x$ and $y$ are small, $k_0\,{\simeq}\,(\pi{-}x)/2$ and
$\Delta\,{\simeq}\,y^2/2$, so $x$ rules the position of the peak,
while $y$ determines its width.

\section{large-$N$ limit of the amplitude}
\label{a.largeN}

The transition amplitude, Eq.~\eqref{e.amplitude}, in the case of odd
$N\,{=}\,2M{-}1$ reads
\begin{equation}
 u(t) = \frac{\Delta(1+\Delta)}{N+3}
      \sum_{m=-M}^M \frac{e^{\imath(\pi m-t\sin q_m)}}{\Delta^2+\tan^2q_m}~,
\end{equation}
where the summation has been made symmetric through the change of
variable $q\,{=}\,\pi/2\,{-}\,k$. The shift equation~\eqref{e.x0.kn}
turns into
\begin{equation}
 \pi\,m = (N{+}3)~q_m+ \pi\varphi_{q_m} ~,
\end{equation}
with
\begin{equation}
 \pi \varphi_q = 2\Big(\tan^{-1}\frac{\tan q}{\Delta}-q\Big)~.
\end{equation}
In the limit $N\,{\to}\,\infty$ one can write the sum as an integral
setting
\begin{equation}
 \frac1{N{+}3}\sum_m
 ~\longrightarrow~ \int\frac{d q}\pi\Big(1+\frac{\pi\varphi'_q}{N{+}3}\Big)
 ~\longrightarrow~ \int\frac{d q}\pi~.
\end{equation}
As we deal within the region of the optimal value of
$\Delta\,{\sim}\,{N^{-1/3}\,{\to}\,{0}}$, we have
\begin{equation}
 u_\infty(t) = \lim_{N\to\infty}
        \Delta\!\!\int_{-\frac\pi2}^{\frac\pi2} \frac{dq}\pi~
        \frac{e^{\imath[(N{+}3)q+\pi\varphi_q{-}t\sin q]}}
             {\Delta^2+\tan^2q}~.
\end{equation}
Writing the arrival time as $t\,{=}\,N{+}3\,{+}\,s$, where $s$ is the
arrival delay, one has then
\begin{equation}
 u_\infty(t) =
  \lim_{t\to\infty}\Delta\int_{-\frac\pi2}^{\frac\pi2} \frac{dq}\pi~
  \frac{e^{\imath[t(q{-}\sin q)-sq+\pi\varphi_q]}}{\Delta^2+\tan^2q}
\end{equation}
The relevant $q$'s are of the order of
$\Delta\,{\sim}\,{N^{-1/3}\to{0}}$, so we change to
$q\,{=}\,\Delta{x}$, with $x$ of the order of 1; keeping the leading
terms for $\Delta\to{0}$,
\begin{eqnarray}
 t\,(q-\sin q) ~~&\longrightarrow&~~ \frac{t\Delta^3}6~x^3 ~,
\\
 \pi~\varphi_q  ~~&\longrightarrow&~~ 2\,\tan^{-1}x ~,
\\
 \frac{\Delta~dq}{\Delta^2+\tan^2q}
 ~~&\longrightarrow&~~ \frac{dx}{1+ x^2} ~,
\end{eqnarray}
and defining the rescaled counterparts of the arrival time
$t\,{\simeq}\,N$ and of the delay $s\,{\sim}\,{N^{1/3}}$,
\begin{equation}
 \tau \equiv \frac{\Delta^3}6~t
~,~~~~~~~
 \sigma \equiv \Delta~s ~;
\label{e.tausigma}
\end{equation}
the final asymptotic expression results
\begin{equation}
 u_\infty(\tau,\sigma) =\int_{-\infty}^{\infty} \frac{dx}\pi~
        \frac{e^{\imath(\tau x^3-\sigma x+2\tan^{-1}x)}}{1+x^2}~,
\label{e.uinfty}
\end{equation}
that can also be rewritten in the form of a simple summation of
phases by introducing the variable $z\,{=}\,\tan^{-1}x$,
\begin{equation}
   u_\infty(\tau,\sigma)
   = \frac2\pi \int_0^{\frac\pi2} dz~
        \cos(\tau \tan^3{z}-\sigma\tan{z}+2z) ~.
\label{e.ucos}
\end{equation}
As in the finite-$N$ case, one has to maximize
$u_\infty(\tau,\sigma)$ by finding the optimal values of $\sigma$ and
$\tau$. For $\tau\,{=}\,0$ it is easy to evaluate
Eq.~\eqref{e.uinfty} analytically,
\begin{equation}
 u_\infty(0,\sigma) =  2\,e^{-\sigma}~\sigma ~;
\end{equation}
it is maximal for $\sigma\,{=}\,1$, giving
$u(0,1)\,{=}\,2e^{-1}\,{\simeq}\,0.736$, to be regarded as a lower
bound to the overall maximum of $u_\infty(\tau,\sigma)$. The overall
maximization has been performed numerically using Eq.~\eqref{e.ucos}.
It turns out that the maximum corresponds to $\sigma\,{=}\,1.2152$
and $\tau\,{=}\,0.02483$, and amounts to
$u_\infty(0.02483,1.2152)\,{=}\,0.84690$, in agreement with the
behavior shown in Fig.~\ref{f.Umax}. The resulting scaling, from
Eq.~\eqref{e.tausigma}, tells that asymptotically
\begin{equation}
 \Delta \simeq  0.530~N^{-1/3}
~,~~~
 s \simeq 2.29~N^{1/3}~.
\label{e.scaling}
\end{equation}

\end{document}